\documentclass[notitlepage,twocolumn]{article}

\makeatletter
\usepackage{paralist}
\usepackage{subcaption}
\usepackage{booktabs}
\usepackage{amsmath}
\usepackage{amssymb}
\usepackage{graphicx}
\usepackage{calc}

\let\@footnotemark@nolink\@footnotemark
\let\@footnotetext@nolink\@footnotetext
\RequirePackage[bookmarksnumbered,unicode]{hyperref}
\RequirePackage{graphicx, xcolor}
\RequirePackage{geometry}
\RequirePackage{manyfoot}
\RequirePackage[tt=false]{libertine}
\RequirePackage{caption, float}
\RequirePackage{comment}
\RequirePackage{textcomp}
\RequirePackage{fancyhdr}
\RequirePackage{xkeyval}
\RequirePackage{microtype}
\RequirePackage{totpages}
\RequirePackage{environ}
\RequirePackage{setspace}
\RequirePackage[numbers]{natbib}

\hypersetup{%
 colorlinks=true,
 linkcolor=blue,
 citecolor=blue,
 urlcolor=blue
}

\newcommand{\para}[1]{\smallskip
\noindent \textbf{#1}} 
   
\makeatother

\title{Benchmark for Complex Answer Retrieval}

\author{
    Federico Nanni\\
    University of Mannheim\\
    fede@informatik.uni-mannheim.de
  \and
    Bhaskar Mitra\\
    Microsoft, UCL\thanks{The author is a part-time PhD student at University College London.}\\
    bmitra@microsoft.com
    \and
    Matt Magnusson\\
    University of New Hampshire\\
    magnusson3@gmail.com
    \and
    Laura Dietz\\
    University of New Hampshire\\
    dietz@cs.unh.edu
}

\date{}
\begin{document}

\maketitle

\begin{abstract}
Retrieving paragraphs to populate a Wikipedia article is a challenging task. The new TREC Complex Answer Retrieval (TREC CAR) track introduces a comprehensive dataset that targets this retrieval scenario. We present early results from a variety of approaches -- from standard information retrieval methods (e.g., tf-idf) to  complex systems that using query expansion using knowledge bases and deep neural networks. The goal is to offer future participants of this track an overview of some promising approaches to tackle this problem.
\end{abstract}

\section{Introduction}


Over the last two decades, research in information retrieval (IR) has developed a variety of approaches for answering queries regarding precise facts -- such as ``Population New York City'', ``Who is Bill de Blasio?'' or ``Neighborhoods in Manhattan'' -- via the identification, extraction, and synthesis of pieces of information from textual sources. However, for more complex queries, such as ``Benefits of immigration for NYC culture'' current systems still rely on presenting the traditional ten blue links to the user as an answer.

To solicit works in this direction, and following the output of the recent SWIRL 2012 workshop on frontiers, challenges, and opportunities for information retrieval report \cite{allan2012frontiers}, a new TREC track on Complex Answer Retrieval\footnote{http://trec-car.cs.unh.edu/} (TREC CAR) for open-domain queries has been recently introduced \cite{dietz2017trec-car-data}.

The task and related dataset are based on the assumption that each Wikipedia page represents a complex topic, with further details under each sections. Accordingly, paragraphs contained in a section such as ``Cultural diversity -- Demographic'' of the page ``Culture of New York City'', offer one aspect of the open-domain query ``Culture of New York City''. The goal of the task is presented as such: given an outline of a page (in the form of the page title and hierarchical section headings), retrieve a ranking of passages for each section. While assessed manually in the future, in this work, a passage is relevant for a section if and only if it is contained in the original article in the corresponding section.

\para{Contribution.} Many different approaches can be applied to this problem. This can address the ways a query can be expanded (using textual or structural information from knowledge bases), the way query and passages can be represented as vectors (e.g., word embedding vectors), and applications of deep neural networks and learning to rank. Given the recent release of the dataset, with this work we intend to support the future participants of the TREC CAR task by studying the performance of a variety of methods---to highlight which of them may be the most promising directions.

Additionally with the publication of this paper, we make our experimentation environment and dataset available\footnote{URL withheld for blind review}, which simulates a noisy candidate generation method for this task.

\para{Outline.} We give an overview of related work in Section 2, describe the data set and our experimentation environment in Section 3 and 4. Section 5 provides details of the approaches. We evaluate empirically in Section 6 before concluding the paper.
\section{Related Work}

A wide variety of approaches are applicable to the TREC CAR problem. In the following, we cover three central ones, namely passage retrieval, query expansion using knowledge bases and the recent advancement in the use of deep neural network models for information retrieval.

\para{Passage Retrieval.} Passage retrieval is often cast as a variation on document retrieval, where the document retrieval model is
applied only to a fragment of the text. The applications include search snippet detection, which aims to summarize the query-relevant parts of a document. 
 Scores under the passage model
can be combined with those from the containing document to improve
performance \cite{bendersky2008passageir} or to include quality
indicators \cite{bendersky2011quality}.  These approaches have been
adapted to retrieve answers for questions \cite{aktolga2011passage}. Passage retrieval models can be extended to combine terms and entity-centric knowledge \cite{blanco10, dietz2015interface}. For certain Wikipedia categories, template of articles can be extracted and automatically populated \cite{sauper2009wikigeneration}. Furthermore, Banerjee and Mitra \cite{banerjee2015wikikreator} found that training a lexical classifier per section heading obtains good results for article construction.

\para{Neural-IR.} With the comeback of neural networks, the IR community is exploring pre-trained word-vector approaches as well as dedicated neural networks for ranking. Pre-trained word- and entity-embeddings are publicly available in the form of word2vec,\footnote{\url{https://code.google.com/archive/p/word2vec/}} GloVe,\footnote{\url{http://nlp.stanford.edu/projects/glove/}} DESM,\footnote{\url{https://www.microsoft.com/en-us/download/details.aspx?id=52597}}  NTLM,\footnote{\url{http://www.zuccon.net/ntlm.html}} wiki2vec,\footnote{\url{https://github.com/idio/wiki2vec}} and RDF2Vec\footnote{\url{http://data.dws.informatik.uni-mannheim.de/rdf2vec/}} vectors. Much of the work in this area focuses on the applications of these shallow distributional models to IR tasks, although recently deeper architectures have also been investigated \cite{mitra2017neural, zhang2016neural}.

\para{Query Expansion with KB.} 
Pseudo relevance feedback \cite{lavrenko2001rm3} is one of the most popular query expansion methods in which frequent words in top documents of a first retrieval run are extracted. This idea is generalized to expansion with multiple sources \cite{bendersky2012multisourceexpansion} based on terms and phrases. 
Recent developments in entity linking algorithms and object retrieval make it feasible to efficiently tap into the rich information provided by KBs \cite{kotov2012tapping,liu2015les, dalton14}, and exploit disambiguation and confidences for query and document representation \cite{hasibi-entitylinking,raviv2016entitylinking}. Further work on entity aspects \cite{reinanda-mining-2015} and effective learning-to-rank approaches for latent entities \cite{xiong2015esdrank} are a promising avenue.


\section{Data Set}

The goal of TREC CAR task is to, given a title and an outline of section headings as a query, retrieve and rank paragraphs from a provided collection \cite{dietz2017trec-car-data}.

The TREC CAR organizers process the English Wikipedia to obtain only the articles with hierarchical section headings (discarding info boxes, images and wrappers). From each article the outline of hierarchical headings are retained, while all paragraphs on the article are split out. All seven million paragraphs from all Wikipedia pages are shuffled to form the \textbf{Paragraph collection}.

\para{Train data.} For 50\% of all articles, both outlines and articles are made available as training data for supervised machine learning as well as a resource for the Rocchio method discussed below.
 
%
%
%
%
%
%

\para{Test200.} The organizers provide a manually selected test collection of outlines from 200 articles. Together, these 200 outlines include approximately 2300 headings, each resulting in a query. This test data is complemented with a ground truth file of which paragraphs are relevant for a given heading, which we base this study on.



\section{Experimentation Environment}
In this paper, we focus on the task of retrieving and ranking paragraphs for each heading of the outline. We consider the 2300 sections in Test200 corpus: First we experiment in a ``safe'' experimentation environment before applying approaches the entire collection of seven million paragraphs. Code and data for a experimentation environment are provided with this publication.

The goal is to simulate a noisy candidate generation method which is guaranteed to include all relevant paragraphs for the heading with a set of nearly-relevant negatives. 

For each heading, we construct a \textbf{train set} by selecting, for every true paragraph under the heading, five paragraphs from different sections of the article, and five paragraphs from a different article.

Similarly we construct a \textbf{test set} that includes all paragraphs from the article, as well as the same amount of paragraphs drawn from other articles. All articles are provided in random order. On average this process yields a mean of 35 paragraphs per section.

\section{Examined Approaches}

The setup of TREC CAR is a bit unusual as all headings of an article are given at once, with the goal of producing a separate ranking for each heading. In this early work, we are breaking each outline into several independent queries, one per heading $h$ as follows: We identify the path from the heading to the root, and concatenate all all headings together with the page title to obtain the query. For example, if $h$ corresponds to heading H2.3.4 the query is the concatenation of H2.3.4, H2.3, H2 and the page title.

Based on these queries, we experiment with different query expansion approaches and vector space representations of queries and paragraphs (tf-idf, GloVe embeddings and RDF2Vec embeddings). We examine BM25, cosine similarity, learning to rank \cite{li2014learning} and a state-of-the-art neural network model \cite{mitra2016learning}.\smallskip

\subsection{Query Expansion Techniques}

We experiment three different query expansion approaches:

\para{Expansion terms (RM1).} Feedback terms are derived using pseudo relevance feedback and the relevance model \cite{lavrenko2001rm3}. Here we use Galago's implementation\footnote{lemurproject.org/galago.php} which is based on a Dirichlet smoothed language model for the feedback run. In the experimental setting, we achieve the best performance expanding the queries with top 10 terms extracted from the top 10 feedback paragraphs.

\para{Expansion entities (ent-RM1).} Another way of expanding queries is by retrieving relevant entities. As for retrieving supporting terms, we derive a set of feedback entities by a search of the index using the heading-query and deriving several entities. In the experimental setting, best performance are achieved using 10 entities.

\para{Paragraph Rocchio.} Inspired by the work of Banerjee and Mitra \cite{banerjee2015wikikreator}, we retrieve other paragraphs, which have an identical heading to our heading-query, from folds 1 to 4 of the collection (omitting the fold where test200 originates from). For example, given a query such as ``Demographic'', regarding the entity United States, we collect supporting paragraphs from the pages of other entities (e.g., United Kingdom), which have as well a a section titled ``Demographic''. Headings are pre-processed with tokenisation, stopword/digit removal and stemming. This way, we can retrieve at least one supporting paragraph for 1/3 of our heading-queries. In the experimental setting, we test expansion using from 1 to 100 supporting passages and we obtain best performance expanding the query with 5 passages.

\subsection{Vector Space Representations}

We study three variations for representing the content in the vector space model:\smallskip

\noindent \textbf{TF-IDF.} 
Representing each word in the vocabulary as its own dimension in the vector space, queries and paragraphs are represented as their TF-IDF vector. We are using the logarithmic L2-normalised variant. We perform stemming as a pre-processing step.

\para{Word Embeddings.} 
Using the pre-trained word embedding GloVE \cite{pennington2014glove} of 300 dimensions, every word $w$ in query or paragraph is represented as a $K$-dimensional vector $\vec{w}$. A vector representation for the whole paragraph $\vec{d}$ (complete query $\vec{q}$) is obtained by a weighted element-wise average of word vectors $\vec{w}$ in the paragraph (query). To give more attention to infrequent word, we use the TF-IDF of each word $w$ to weights.
\begin{equation*}
\vec{d} = \frac{1}{\vert d \vert} \sum_{w\in d} \text{TF-IDF}(w) \cdot \vec{w}
\end{equation*} 
\para{Entity Embeddings.} Queries and paragraphs are represented as their mentioned DBpedia entities, using the entity linker TagMe \cite{ferragina2010tagme} (with default parameters). Next, we obtain latent vector representations $\vec{e}$ of each linked entity $e$ using pre-computed \textbf{RDF2Vec} 300d entity embeddings \cite{ristoski2016rdf2vec}. 
Vector representations of paragraphs $\vec{d}$ (queries $\vec{q}$) are computed by a weighted element-wise average of entity vectors $\vec{e}$. By casting a paragraph as a bag-of-links we adapt TF-IDF to entity links (link statistics taken from DBpedia 2015-04 \cite{auer2007dbpedia}:
\begin{equation*}
\vec{d} = \frac{1}{\vert \left\{ e\in d \right\} \vert} \sum_{e\in d} \text{TF-IDF}(e) \cdot \vec{e}
\end{equation*} 

\subsection{Ranking Approaches} 

We test four different ranking approaches:

\para{Okapi BM25.} Results are ranked using Okapi BM25 with $k_1$=1.2 and $b$=0.75, using the implementation of Lucene 6.4.1. Porter stemming and stopword removal was applied to paragraphs and queries.

\para{Cosine Similarity.} Paragraphs are ranked by cosine similarity (\textbf{cs}) between respective vector representations of the query and the paragraph.

\para{Learning to Rank.} We combine the ranking scores of different baselines with supervised machine learning in a learning-to-rank setting, for producing a final ranking of relevant paragraphs. We use RankLib \footnote{lemurproject.org/ranklib.php} with 5-fold cross validation using a linear model optimized for MAP, trained with coordinate ascent.

\para{Deep Neural Network.} The Duet model is a state-of-the-art deep neural network (DNN) recently proposed by \citet{mitra2016learning} for ad-hoc retrieval. The Duet architecture learns to model query-paragraph relevance by jointly learning good representations of the query and the paragraph text for matching, as well as by learning to identify good patterns of exact matches of the query terms in the paragraph text. We use the Duet implementation available publicly\footnote{\url{https://github.com/bmitra-msft/NDRM/blob/master/notebooks/Duet.ipynb}} under the MIT license for our experiments.

Training on folds 1 to 4 of the collection, we only consider the first ten words for the query and the first 100 words for the passage as inputs. We use 64 hidden units in the different layers of the network, as opposed to 300 in the original paper, to reduce the total number of learnable parameters of the model. We trained the model for 32 epochs with a learning rate of 0.001 which was picked based on a subset of the training data. Each epoch was trained over 1024 minibatches, and each minibatch contained 1024 samples. Each training sample was a triplet consisting of a query, a positive passage, and a negative passage. The training time was limited to 60 hours.
\begin{table}
\caption{\label{table1} Results on experimentation environment.}
\begin{center}

{
\centering
\small{
\begin{tabular}{ lcccc }
\toprule 
  & \textbf{MAP}  & \textbf{R-Prec} & \textbf{MRR} \tabularnewline
\midrule 
\textbf{{bm25} } &  &  & \tabularnewline
query only  & 0.320  & 0.232  & 0.409 \tabularnewline
\midrule 
\textbf{{tf-idf (cs)} } &  &  & \tabularnewline
query only  & 0.350  & 0.212  & 0.383 \tabularnewline
query + RM1  & 0.324  & 0.205  & 0.384 \tabularnewline
query + Rocchio  & 0.400  & 0.286  & 0.466 \tabularnewline
\midrule 
\textbf{{GloVe (cs)} } &  &  & \tabularnewline
query only  & 0.329  & 0.210  & 0.387 \tabularnewline
query + RM1  & 0.289  & 0.177  & 0.339 \tabularnewline
query + Rocchio  & 0.349  & 0.236  & 0.410 \tabularnewline
\midrule 
\textbf{{RDF2Vec (cs)} } &  &  & \tabularnewline
entity-query only  & 0.313  & 0.200  & 0.369 \tabularnewline
ent-query + ent-RM1  & 0.320 & 0.208  & 0.377 \tabularnewline
ent-query + ent-Rocchio  & 0.316 & 0.206  & 0.375 \tabularnewline
\midrule 
\textbf{{Learning to Rank} } &  &  & \tabularnewline
all (cs) scores  & 0.412 & 0.290  & 0.475 \tabularnewline
\midrule 
\textbf{{Duet model} } &  &  & \tabularnewline
query only  & \textbf{{0.470}} & \textbf{{0.359} } & \textbf{{0.553} }\tabularnewline
\bottomrule
\end{tabular}
}
}

\end{center}

\end{table}

\section{Evaluation}

We present evaluation results both on the experimentation environment and on the entire paragraph collection.

\subsection{Experimentation Environment}

The experimentation environment (Section 4) provides a ``safe environment'' by simulating a noisy candidate method for each section. Results are presented in table \ref{table1}. The approach \emph{bm25 query only} sets the baseline of our work. 

Not all query expansion approaches and vector space representation methods improve over this baseline. This is particularly true for query expansion with terms or entities (through RM3 = query + RM1) as well as RDF2Vec embeddings. On the contrary, the most promising results among the methods which employ cosine similarity as a ranking function, are obtained when the query is expanded with Rocchio vectors trained on paragraphs from sections with the same heading. This finding reconfirms the results of previous work on the automatic generation of Wikipedia articles based its structural information \cite{banerjee2015wikikreator}. The results show that common traits between Wikipedia sections with the same heading are better captured using the tf-idf word vector than through word- and entity- embedding vectors suggest that a possible improvement over these baselines could by obtained by training embeddings for this task.

\begin{figure}
\begin{centering}
\includegraphics[width=3in]{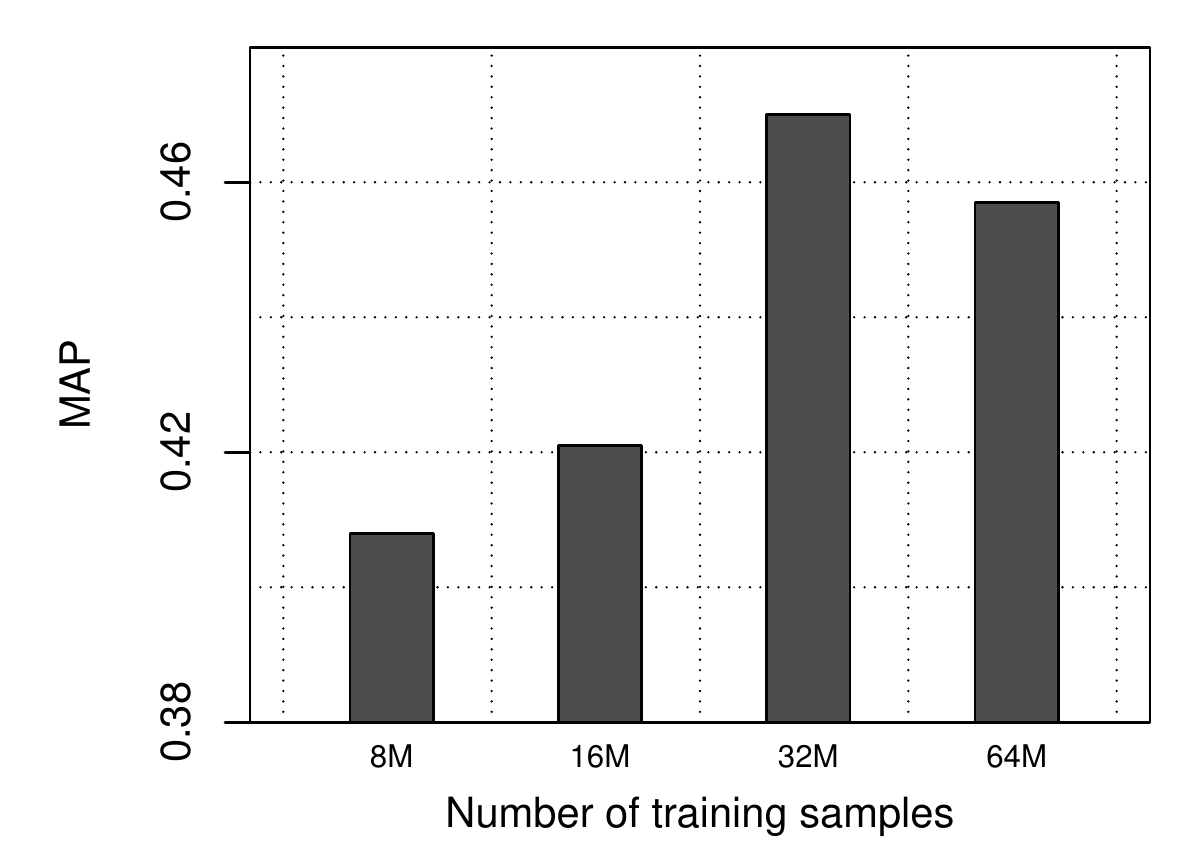}
\par\end{centering}
\caption{Effect of training data size on the  performance of the Duet model. Training on four folds, reporting MAP on holdout fold.}
\label{fig:trainingdata}
\end{figure}

In comparison to these unsupervised retrieval models, both supervised Learning to Rank and the Neural Duet model out-perform all previously described baselines. In particular, the Duet model yields a substantial improvement over all presented approaches, showing the potential of neural-IR for the task. It is important to remark that neural deep models take days to train even on a GPU. In addition they are data-hungry, with performances improving significantly with more training data as shown in Figure \ref{fig:trainingdata}.


\begin{table}
\caption{\label{table2} Results of the initial candidate selection.}
\begin{center}
{
\centering
\small{
\begin{tabular}{ lcccc }
\toprule 
{\small{} } & \textbf{\small{}MAP}{\small{} } & \textbf{\small{}R-Prec} & \textbf{\small{}MRR}{\small{} }  \tabularnewline
\midrule 
\textbf{\small{}{Heading Retrieval - bm25} } &  &  &   \tabularnewline
{\small{}query only } & {\small{}\textbf{0.150} } & {\small{}\textbf{0.118} } & {\small{}\textbf{0.216} }  \tabularnewline
\midrule 
\textbf{\small{}{Heading Retrieval - tf-idf (cs)} } &  &  &   \tabularnewline
{\small{}query only } & \small{}0.035{\small{} } & \small{}0.025{\small{} } & \small{}0.053{\small{} }  \tabularnewline
{\small{}query + Rocchio } & {\small{}0.029} & {\small{}0.020 } & {\small{}0.041 }  \tabularnewline
\bottomrule
\end{tabular}
}
}
\end{center}
\end{table}

\begin{table*}[t]
\caption{\label{table3} Results on the Paragraph Collection after candidate method. $^\blacktriangledown$  Worse according to paired-t-test with $\alpha=5\%$.}
\begin{center}

{ \centering {\small{}{ }%
\begin{tabular}{     lcccccccccc }
\toprule 
& & &  &  & w/o bm25 \tabularnewline
{\small{} } & \textbf{\small{}MAP}{\small{} } & \textbf{\small{}R-Prec} & \textbf{\small{}MRR}{\small{} } &  & \textbf{\small{}MAP}{\small{} }  \tabularnewline
\midrule 
bm25 candidate & 0.150$^\blacktriangledown$  & 0.118 & 0.216 &  & - \tabularnewline
\midrule 
\textbf{\small{}{tf-idf (cs)} } &  &  &  &  & \tabularnewline
{\small{}bm25 + query + Rocchio } & 0.140$^\blacktriangledown$  & 0.113 & 0.207 &  & {\small{}0.085} \tabularnewline
\midrule 
\textbf{\small{}{GloVe (cs)} } &  &  &  &  &   \tabularnewline
{\small{}bm25 + query + Rocchio } & 0.150$^\blacktriangledown$  & 0.120 & 0.218 &  & {\small{}0.082} \tabularnewline
\midrule 
\textbf{\small{}{Duet model} } &  &  &  &  &   \tabularnewline
{\small{}bm25 + query only } & \textbf{\small{}{0.161}} & \textbf{\small{}{0.130} } & \textbf{\small{}{0.229} } &  & \textbf{\small{}{0.093}}  \tabularnewline
\bottomrule
\end{tabular}{\small{}} } }{\small \par}
\end{center}
\end{table*}

\subsection{Experiments on Paragraph Collection}

As a second step, we conduct experiments on the entire paragraph collection. 

\noindent \textbf{Candidate Selection.} Many of the previously presented methods, such as the Duet model, require a candidate generating method. We test three candidate methods: bm25, tf-idf, and tf-idf with rocchio expansion. For each query, the methods produce a candidate set of 100 paragraphs.
%
%
 The results are presented in Table \ref{table2}. While this is a challenging task, encouraging performance are obtained with the use of \emph{bm25 query only}. However as half of the queries do not have any relevant paragraphs in the candidate set, the theoretically achievable MRR is 0.50.
  

As a final experiment, we use \emph{bm25 query only} as a candidate method to obtain 100 paragraphs per query.  The three of the most-promising systems presented above are re-evaluated, using learning to rank to combine the scores for bm25, query, and Rocchio expansion with 5-fold cross validation. 

The combination of bm25 score and duet model is significantly outperforming all other methods, demonstrating the strength of the candidate method. If the combination with the bm25 score is left out, all methods are significantly loosing in performance (see last column in Table \ref{table3}).


\section{Conclusions}

In this paper, we present the performance of a variety of approaches, from established baselines to more advanced systems, in the context of the new TREC-CAR track on Complex Answer Retrieval.  Our results show the effects of different query expansion and embedding approaches in comparison to learning to rank and neural ranking models. The public availability of this new dataset, gave us the opportunity to offer a benchmark on the most promising approaches for tackling this problem to future participants of the track and the IR community.

\bibliographystyle{ACM-Reference-Format}
\bibliography{sigproc}


\begin{thebibliography}{00}


\ifx \showCODEN    \undefined \def \showCODEN     #1{\unskip}     \fi
\ifx \showDOI      \undefined \def \showDOI       #1{{\tt DOI:}\penalty0{#1}\ }
  \fi
\ifx \showISBNx    \undefined \def \showISBNx     #1{\unskip}     \fi
\ifx \showISBNxiii \undefined \def \showISBNxiii  #1{\unskip}     \fi
\ifx \showISSN     \undefined \def \showISSN      #1{\unskip}     \fi
\ifx \showLCCN     \undefined \def \showLCCN      #1{\unskip}     \fi
\ifx \shownote     \undefined \def \shownote      #1{#1}          \fi
\ifx \showarticletitle \undefined \def \showarticletitle #1{#1}   \fi
\ifx \showURL      \undefined \def \showURL       #1{#1}          \fi
\providecommand\bibfield[2]{#2}
\providecommand\bibinfo[2]{#2}
\providecommand\natexlab[1]{#1}
\providecommand\showeprint[2][]{arXiv:#2}

\bibitem[\protect\citeauthoryear{Aktolga, Allan, and Smith}{Aktolga
  et~al\mbox{.}}{2011}]%
        {aktolga2011passage}
\bibfield{author}{\bibinfo{person}{Elif Aktolga}, \bibinfo{person}{James
  Allan}, {and} \bibinfo{person}{David~A. Smith}.}
  \bibinfo{year}{2011}\natexlab{}.
\newblock \showarticletitle{Passage reranking for question answering using
  syntactic structures and answer types}.
\newblock In \bibinfo{booktitle}{{\em Advances in Information Retrieval}}.
  \bibinfo{publisher}{Springer}.
\newblock


\bibitem[\protect\citeauthoryear{Allan, Croft, Moffat, and Sanderson}{Allan
  et~al\mbox{.}}{2012}]%
        {allan2012frontiers}
\bibfield{author}{\bibinfo{person}{James Allan}, \bibinfo{person}{Bruce Croft},
  \bibinfo{person}{Alistair Moffat}, {and} \bibinfo{person}{Mark Sanderson}.}
  \bibinfo{year}{2012}\natexlab{}.
\newblock \showarticletitle{Frontiers, challenges, and opportunities for
  information retrieval: Report from SWIRL 2012 the second strategic workshop
  on information retrieval in Lorne}. In \bibinfo{booktitle}{{\em ACM SIGIR
  Forum}}, Vol.~\bibinfo{volume}{46}. ACM, \bibinfo{pages}{2--32}.
\newblock


\bibitem[\protect\citeauthoryear{Auer, Bizer, Kobilarov, Lehmann, Cyganiak, and
  Ives}{Auer et~al\mbox{.}}{2007}]%
        {auer2007dbpedia}
\bibfield{author}{\bibinfo{person}{S{\"o}ren Auer}, \bibinfo{person}{Christian
  Bizer}, \bibinfo{person}{Georgi Kobilarov}, \bibinfo{person}{Jens Lehmann},
  \bibinfo{person}{Richard Cyganiak}, {and} \bibinfo{person}{Zachary Ives}.}
  \bibinfo{year}{2007}\natexlab{}.
\newblock \bibinfo{booktitle}{{\em Dbpedia: A nucleus for a web of open data}}.
\newblock \bibinfo{publisher}{Springer}.
\newblock


\bibitem[\protect\citeauthoryear{Banerjee and Mitra}{Banerjee and
  Mitra}{2015}]%
        {banerjee2015wikikreator}
\bibfield{author}{\bibinfo{person}{Siddhartha Banerjee} {and}
  \bibinfo{person}{Prasenjit Mitra}.} \bibinfo{year}{2015}\natexlab{}.
\newblock \showarticletitle{WikiKreator: Improving Wikipedia Stubs
  Automatically.}. In \bibinfo{booktitle}{{\em ACL (1)}}.
  \bibinfo{pages}{867--877}.
\newblock


\bibitem[\protect\citeauthoryear{Bendersky, Croft, and Diao}{Bendersky
  et~al\mbox{.}}{2011}]%
        {bendersky2011quality}
\bibfield{author}{\bibinfo{person}{Michael Bendersky}, \bibinfo{person}{W~Bruce
  Croft}, {and} \bibinfo{person}{Yanlei Diao}.}
  \bibinfo{year}{2011}\natexlab{}.
\newblock \showarticletitle{Quality-biased ranking of web documents}. In
  \bibinfo{booktitle}{{\em WSDM}}.
\newblock


\bibitem[\protect\citeauthoryear{Bendersky and Kurland}{Bendersky and
  Kurland}{2008}]%
        {bendersky2008passageir}
\bibfield{author}{\bibinfo{person}{Michael Bendersky} {and}
  \bibinfo{person}{Oren Kurland}.} \bibinfo{year}{2008}\natexlab{}.
\newblock \showarticletitle{Utilizing passage-based language models for
  document retrieval}.
\newblock In \bibinfo{booktitle}{{\em Advances in Information Retrieval}}.
  \bibinfo{publisher}{Springer}.
\newblock


\bibitem[\protect\citeauthoryear{Bendersky, Metzler, and Croft}{Bendersky
  et~al\mbox{.}}{2012}]%
        {bendersky2012multisourceexpansion}
\bibfield{author}{\bibinfo{person}{Michael Bendersky}, \bibinfo{person}{Donald
  Metzler}, {and} \bibinfo{person}{W~Bruce Croft}.}
  \bibinfo{year}{2012}\natexlab{}.
\newblock \showarticletitle{Effective query formulation with multiple
  information sources}. In \bibinfo{booktitle}{{\em WSDM}}.
\newblock


\bibitem[\protect\citeauthoryear{Blanco and Zaragoza}{Blanco and
  Zaragoza}{2010}]%
        {blanco10}
\bibfield{author}{\bibinfo{person}{Roi Blanco} {and} \bibinfo{person}{Hugo
  Zaragoza}.} \bibinfo{year}{2010}\natexlab{}.
\newblock \showarticletitle{{F}inding support sentences for entities}. In
  \bibinfo{booktitle}{{\em SIGIR}}.
\newblock


\bibitem[\protect\citeauthoryear{Dalton, Dietz, and Allan}{Dalton
  et~al\mbox{.}}{2014}]%
        {dalton14}
\bibfield{author}{\bibinfo{person}{Jeffrey Dalton}, \bibinfo{person}{Laura
  Dietz}, {and} \bibinfo{person}{James Allan}.}
  \bibinfo{year}{2014}\natexlab{}.
\newblock \showarticletitle{Entity Query Feature Expansion Using Knowledge Base
  Links}. In \bibinfo{booktitle}{{\em SIGIR}}.
\newblock


\bibitem[\protect\citeauthoryear{Dietz and Gamari}{Dietz and Gamari}{2017}]%
        {dietz2017trec-car-data}
\bibfield{author}{\bibinfo{person}{Laura Dietz} {and} \bibinfo{person}{Ben
  Gamari}.} \bibinfo{year}{2017}\natexlab{}.
\newblock \bibinfo{title}{{TREC CAR}: A Data Set for {C}omplex {A}nswer
  {R}etrieval}.
\newblock \bibinfo{howpublished}{Version 1.4}.   (\bibinfo{year}{2017}).
\newblock
\showURL{%
\url{http://trec-car.cs.unh.edu}}


\bibitem[\protect\citeauthoryear{Dietz and Schuhmacher}{Dietz and
  Schuhmacher}{2015}]%
        {dietz2015interface}
\bibfield{author}{\bibinfo{person}{Laura Dietz} {and} \bibinfo{person}{Michael
  Schuhmacher}.} \bibinfo{year}{2015}\natexlab{}.
\newblock \showarticletitle{An interface sketch for queripidia: Query-driven
  knowledge portfolios from the web}. In \bibinfo{booktitle}{{\em Proc.
  Workshop on Exploiting Semantic Annotations in IR}}. ACM,
  \bibinfo{pages}{43--46}.
\newblock


\bibitem[\protect\citeauthoryear{Ferragina and Scaiella}{Ferragina and
  Scaiella}{2010}]%
        {ferragina2010tagme}
\bibfield{author}{\bibinfo{person}{Paolo Ferragina} {and} \bibinfo{person}{Ugo
  Scaiella}.} \bibinfo{year}{2010}\natexlab{}.
\newblock \showarticletitle{Tagme: on-the-fly annotation of short text
  fragments (by wikipedia entities)}. In \bibinfo{booktitle}{{\em CIKM}}. ACM.
\newblock


\bibitem[\protect\citeauthoryear{Hasibi, Balog, and Bratsberg}{Hasibi
  et~al\mbox{.}}{2016}]%
        {hasibi-entitylinking}
\bibfield{author}{\bibinfo{person}{Faegheh Hasibi}, \bibinfo{person}{Krisztian
  Balog}, {and} \bibinfo{person}{Svein~Erik Bratsberg}.}
  \bibinfo{year}{2016}\natexlab{}.
\newblock \showarticletitle{Exploiting Entity Linking in Queries for Entity
  Retrieval}. In \bibinfo{booktitle}{{\em ICTIR}}. \bibinfo{pages}{209--218}.
\newblock


\bibitem[\protect\citeauthoryear{Kotov and Zhai}{Kotov and Zhai}{2012}]%
        {kotov2012tapping}
\bibfield{author}{\bibinfo{person}{Alexander Kotov} {and}
  \bibinfo{person}{ChengXiang Zhai}.} \bibinfo{year}{2012}\natexlab{}.
\newblock \showarticletitle{Tapping into knowledge base for concept feedback:
  leveraging conceptnet to improve search results for difficult queries}. In
  \bibinfo{booktitle}{{\em WSDM}}.
\newblock


\bibitem[\protect\citeauthoryear{Lavrenko and Croft}{Lavrenko and
  Croft}{2001}]%
        {lavrenko2001rm3}
\bibfield{author}{\bibinfo{person}{Victor Lavrenko} {and}
  \bibinfo{person}{W~Bruce Croft}.} \bibinfo{year}{2001}\natexlab{}.
\newblock \showarticletitle{Relevance based language models}. In
  \bibinfo{booktitle}{{\em SIGIR}}.
\newblock


\bibitem[\protect\citeauthoryear{Li}{Li}{2014}]%
        {li2014learning}
\bibfield{author}{\bibinfo{person}{Hang Li}.} \bibinfo{year}{2014}\natexlab{}.
\newblock \showarticletitle{Learning to rank for information retrieval and
  natural language processing}.
\newblock \bibinfo{journal}{{\em Synthesis Lectures on Human Language
  Technologies\/}} \bibinfo{volume}{7}, \bibinfo{number}{3}
  (\bibinfo{year}{2014}).
\newblock


\bibitem[\protect\citeauthoryear{Liu and Fang}{Liu and Fang}{2015}]%
        {liu2015les}
\bibfield{author}{\bibinfo{person}{Xitong Liu} {and} \bibinfo{person}{Hui
  Fang}.} \bibinfo{year}{2015}\natexlab{}.
\newblock \showarticletitle{Latent entity space: a novel retrieval approach for
  entity-bearing queries}.
\newblock \bibinfo{journal}{{\em Information Retrieval Journal\/}}
  \bibinfo{volume}{18}, \bibinfo{number}{6} (\bibinfo{year}{2015}).
\newblock


\bibitem[\protect\citeauthoryear{Mitra and Craswell}{Mitra and
  Craswell}{2017}]%
        {mitra2017neural}
\bibfield{author}{\bibinfo{person}{Bhaskar Mitra} {and} \bibinfo{person}{Nick
  Craswell}.} \bibinfo{year}{2017}\natexlab{}.
\newblock \showarticletitle{Neural Models for Information Retrieval}.
\newblock \bibinfo{journal}{{\em arXiv preprint arXiv:1705.01509\/}}
  (\bibinfo{year}{2017}).
\newblock


\bibitem[\protect\citeauthoryear{Mitra, Diaz, and Craswell}{Mitra
  et~al\mbox{.}}{2017}]%
        {mitra2016learning}
\bibfield{author}{\bibinfo{person}{Bhaskar Mitra}, \bibinfo{person}{Fernando
  Diaz}, {and} \bibinfo{person}{Nick Craswell}.}
  \bibinfo{year}{2017}\natexlab{}.
\newblock \showarticletitle{Learning to Match Using Local and Distributed
  Representations of Text for Web Search}. In \bibinfo{booktitle}{{\em www}}.
\newblock


\bibitem[\protect\citeauthoryear{Pennington, Socher, and Manning}{Pennington
  et~al\mbox{.}}{2014}]%
        {pennington2014glove}
\bibfield{author}{\bibinfo{person}{Jeffrey Pennington},
  \bibinfo{person}{Richard Socher}, {and} \bibinfo{person}{Christopher~D
  Manning}.} \bibinfo{year}{2014}\natexlab{}.
\newblock \showarticletitle{Glove: Global Vectors for Word Representation.}. In
  \bibinfo{booktitle}{{\em EMNLP}}, Vol.~\bibinfo{volume}{14}.
\newblock


\bibitem[\protect\citeauthoryear{Raviv, Kurland, and Carmel}{Raviv
  et~al\mbox{.}}{2016}]%
        {raviv2016entitylinking}
\bibfield{author}{\bibinfo{person}{Hadas Raviv}, \bibinfo{person}{Oren
  Kurland}, {and} \bibinfo{person}{David Carmel}.}
  \bibinfo{year}{2016}\natexlab{}.
\newblock \showarticletitle{Document Retrieval Using Entity-Based Language
  Models}. In \bibinfo{booktitle}{{\em SIGIR}}.
\newblock


\bibitem[\protect\citeauthoryear{Reinanda, Meij, and de~Rijke}{Reinanda
  et~al\mbox{.}}{2015}]%
        {reinanda-mining-2015}
\bibfield{author}{\bibinfo{person}{Ridho Reinanda}, \bibinfo{person}{Edgar
  Meij}, {and} \bibinfo{person}{Maarten de Rijke}.}
  \bibinfo{year}{2015}\natexlab{}.
\newblock \showarticletitle{Mining, ranking and recommending entity aspects}.
  In \bibinfo{booktitle}{{\em SIGIR}}.
\newblock


\bibitem[\protect\citeauthoryear{Ristoski and Paulheim}{Ristoski and
  Paulheim}{2016}]%
        {ristoski2016rdf2vec}
\bibfield{author}{\bibinfo{person}{Petar Ristoski} {and} \bibinfo{person}{Heiko
  Paulheim}.} \bibinfo{year}{2016}\natexlab{}.
\newblock \showarticletitle{Rdf2vec: Rdf graph embeddings for data mining}. In
  \bibinfo{booktitle}{{\em ISWC}}. Springer.
\newblock


\bibitem[\protect\citeauthoryear{Sauper and Barzilay}{Sauper and
  Barzilay}{2009}]%
        {sauper2009wikigeneration}
\bibfield{author}{\bibinfo{person}{Christina Sauper} {and}
  \bibinfo{person}{Regina Barzilay}.} \bibinfo{year}{2009}\natexlab{}.
\newblock \showarticletitle{Automatically generating {W}ikipedia articles: A
  structure-aware approach}. In \bibinfo{booktitle}{{\em IJCNLP}}.
\newblock


\bibitem[\protect\citeauthoryear{Xiong and Callan}{Xiong and Callan}{2015}]%
        {xiong2015esdrank}
\bibfield{author}{\bibinfo{person}{Chenyan Xiong} {and} \bibinfo{person}{Jamie
  Callan}.} \bibinfo{year}{2015}\natexlab{}.
\newblock \showarticletitle{Esdrank: Connecting query and documents through
  external semi-structured data}. In \bibinfo{booktitle}{{\em CIKM}}.
\newblock


\bibitem[\protect\citeauthoryear{Zhang, Rahman, Braylan, Dang, Chang, Kim,
  McNamara, Angert, Banner, Khetan, et~al\mbox{.}}{Zhang et~al\mbox{.}}{2016}]%
        {zhang2016neural}
\bibfield{author}{\bibinfo{person}{Ye Zhang}, \bibinfo{person}{Md~Mustafizur
  Rahman}, \bibinfo{person}{Alex Braylan}, \bibinfo{person}{Brandon Dang},
  \bibinfo{person}{Heng-Lu Chang}, \bibinfo{person}{Henna Kim},
  \bibinfo{person}{Quinten McNamara}, \bibinfo{person}{Aaron Angert},
  \bibinfo{person}{Edward Banner}, \bibinfo{person}{Vivek Khetan}, {and}
  \bibinfo{person}{others}.} \bibinfo{year}{2016}\natexlab{}.
\newblock \showarticletitle{Neural Information Retrieval: A Literature Review}.
\newblock \bibinfo{journal}{{\em arXiv preprint arXiv:1611.06792\/}}
  (\bibinfo{year}{2016}).
\newblock


\end{thebibliography}

\end{document}